\documentclass[nofootinbib,notitlepage,superscriptaddress,10pt,aps,pra,twocolumn]{revtex4-2}

\usepackage{graphicx}  
\usepackage{subfigure}
\usepackage{multirow}
\usepackage{ulem}
\usepackage{bm}       
\usepackage{amsfonts}  
\usepackage{amsmath}   
\usepackage{amssymb} 
\usepackage{float}  
\usepackage{cancel}
\usepackage{hyperref}  
\hypersetup{
   pdftitle={Time Dilation in DSR},
   colorlinks=true,
   linkcolor=black,
   urlcolor=blue
}

\begin{document}

\title{DSR-relativistic spacetime picture\\ and the phenomenology of Planck-scale-modified time dilation}

\author{Giovanni Amelino-Camelia}
\affiliation{Dipartimento di Fisica Ettore Pancini, Università di Napoli “Federico II”,
Complesso Univ.\ Monte S.\ Angelo, I-80126 Napoli, Italy}
\affiliation{INFN, Sezione di Napoli, Complesso Univ.\ Monte S.\ Angelo, I-80126 Napoli, Italy}
\author{Giulia Gubitosi}
\affiliation{Dipartimento di Fisica Ettore Pancini, Università di Napoli “Federico II”,
Complesso Univ.\ Monte S.\ Angelo, I-80126 Napoli, Italy}
\affiliation{INFN, Sezione di Napoli, Complesso Univ.\ Monte S.\ Angelo, I-80126 Napoli, Italy}
\author{Pietro Pellecchia}
\affiliation{Dipartimento di Fisica Ettore Pancini, Università di Napoli “Federico II”,
Complesso Univ.\ Monte S.\ Angelo, I-80126 Napoli, Italy}
\affiliation{INFN, Sezione di Napoli, Complesso Univ.\ Monte S.\ Angelo, I-80126 Napoli, Italy}
\author{Marco Refuto}
\affiliation{Dipartimento di Fisica Ettore Pancini, Università di Napoli “Federico II”,
Complesso Univ.\ Monte S.\ Angelo, I-80126 Napoli, Italy}
\author{Giacomo Rosati}
\affiliation{Dipartimento di Matematica, Università di Cagliari, via Ospedale 72, 09124 Cagliari, Italy}
\affiliation{INFN, Sezione di Cagliari, Cittadella Universitaria, 09042 Monserrato, Italy}

\begin{abstract}
The most active area of research in quantum-gravity phenomenology investigates the possibility of  Planck-scale-modified dispersion relations, focusing mainly on two alternative scenarios: the “LIV" scenario, characterized by a specific mechanism of breakdown of relativistic symmetries, and the “DSR" scenario, which preserves overall relativistic invariance but with deformed laws of relativistic transformation.
Two recent studies of modified dispersion relations, one relying on Finsler geometry and one based on heuristic reasoning, raised the possibility of potentially observable effects for time dilation and argued that this might apply also to the LIV and DSR scenarios.
We observe that  the description of Lorentz transformations in the LIV scenario is such that  time dilation cannot be modified. The DSR scenario allows for modifications of time dilation, and establishing their magnitude required us to obtain novel results on the effects of finite DSR boosts in the spacetime sector, with results showing in particular that the modification of time dilation is too small for experimental testing. 
\end{abstract}

\maketitle

\section{Preliminaries on time dilation and dispersion relation}\label{sec:preliminaries}
The young research area of quantum-gravity phenomenology is exploring several avenues for testing candidate effects of quantum gravity (see, {\it e.g.}, Refs.~\cite{gacLivingREview,COSTreview}). The most studied opportunity concerns the possibility that quantum-gravity effects could modify the dispersion relation through a correction term which is cubic in energy-momentum:
\begin{equation}\label{GAC1}
    M^2\approx   \epsilon^2-\vec{p}^{\, 2}-\ell \epsilon \vec{p}^{\, 2},
\end{equation}
where $M$ is the particle's mass, 
$\epsilon$ and $\vec{p}$ denote, respectively, its energy and spatial momentum, and $\ell$ is a characteristic length scale usually assumed to be of the order of the Planck length, up to a numerical factor of order 1 and a possible sign difference~\cite{gacLivingREview,COSTreview}.

The main motivation for such dispersion-relation studies originates from attempts to model quantum properties of spacetime in terms of noncommutativity (see,\ {\it e.g.},\ Refs.~\cite{GACandMajid,DimitrijevicJHEP2011}) or discreteness (see,\ {\it e.g.},\ Refs.~\cite{urrutia, hanno}.
Phenomenologically it is emerging that a key aspect concerns the fate of relativistic symmetries when the dispersion relation is modified. 
A much studied scenario, usually identified by the acronym LIV (Lorentz Invariance Violation), assumes that \eqref{GAC1} is a law that breaks relativistic symmetries (see, {\it e.g.}, Refs.~\cite{gacNATURE1998,piranJCAP2008}): the laws of transformation among observers are still governed by special relativity, resulting in a picture such that \eqref{GAC1} cannot hold for all observers. 
The other much studied scenario, usually identified by the acronym DSR (Doubly-Special or Deformed-Special Relativity), assumes that \eqref{GAC1} is an observer-independent law (see, {\it e.g.}, Refs.~\cite{gacIJMPD2002,gacPLB2001,kowalskiPLB2002,magueijosmolinPRD2003}): this scenario is still fully relativistic but with deformed laws of transformation among observers, with the deformation of course being tailored to render \eqref{GAC1} invariant.\vspace{2mm}

The study we are here reporting is aimed at investigating the possible relevance for the LIV and the DSR scenarios of a new avenue for quantum-gravity phenomenology, proposed in the recent Refs.~\cite{heuristic,fins2,fins1}, that is the implications of Eq.\eqref{GAC1} for time dilation. 
The proposal put forward in Ref.~\cite{heuristic} adopted a heuristic approach such that \eqref{GAC1} was rewritten as $M_{\text{eff}}^2 \approx \epsilon^2 -\vec{p}^{\, 2}$, reabsorbing the cubic correction term into a momentum-dependent redefinition of the mass, and it was then assumed that in a LIV scenario the “gamma factor" for time dilation should be given by $\epsilon/M_{\text{eff}}$.
The end result of Ref.~\cite{heuristic} is that for a high-energy particle ($\epsilon \gg M$) 
the time-dilation factor would be given by
\begin{equation}\label{GAC2}
    \gamma_{dil}(\epsilon) \approx \frac{\epsilon}{M}\left[1 -\ell\frac{\epsilon^3}{2  M^2}\right].
\end{equation}
The surprising aspect of this proposed $\gamma_{dil}(\epsilon)$ is that the correction term is governed by the dimensionless quantity $\ell \epsilon^3/M^2$, while the starting point was a modification of the dispersion relation governed by the dimensionless quantity $\ell \epsilon$.
While $\ell \epsilon$ remains small at all subplanckian energies ($\epsilon \ll \ell^{-1}$), there is clearly a subplankian range of energies ($M \ll \epsilon \ll \ell^{-1}$) such that $\ell \epsilon^3/M^2$ gets  large,  and this could affect in particular high-energy atmospheric muons in a potentially observable way~\cite{heuristic}.

It is not our intention to discourage these phenomenological studies, but in order to properly frame them within the wider picture of quantum-gravity phenomenology we must stress that 
the $\gamma_{dil}$ of \eqref{GAC2} could not possibly be the time dilation factor in the LIV scenario, contrary to what was conjectured in Ref.~\cite{heuristic}. 
This is simply because, as stressed above, the LIV scenario
assumes unmodified laws of relativistic transformation, and therefore automatically predicts an unmodified time dilation.

Intriguingly, a modification of time dilation structurally analogous to that of \eqref{GAC2}
was motivated independently by the analysis based on Finsler geometry reported in Refs.~\cite{fins2, fins1}.
Finsler geometry is being considered as one of the candidates
for an effective  description of the quantum properties of spacetime, and it is known (see, {\it e.g.}, Refs.~\cite{LiberatiSindoni2007,gacGiuliaLiberati}) to be suitable for encoding modifications of the dispersion relation of type \eqref{GAC1}. 
The implementation of Finsler geometry advocated in Refs.~\cite{fins2, fins1} hosts the dispersion relation \eqref{GAC1}  and gives a time dilation factor which, for high-energy particles, is essentially the same as \eqref{GAC2} but with opposite sign of the correction term.

While this is an intrinsically interesting Finsler-geometry result, we are here concerned with the conjecture formulated in Refs.~\cite{fins2,fins1} that the result could be applicable to the DSR scenario. 
That conjecture was based on the fact that some implementations of Finsler geometry are known to be DSR relativistic; however, we must stress that the specific implementation of Finsler geometry advocated in Refs.~\cite{fins2,fins1} is surely not DSR relativistic: as established in Refs.~\cite{mignemiFinslerNoDSR,gacGiuliaLiberati}, while the action based on the Finsler line element leads to a DSR-relativistic description of the particle kinematics, the Finsler line element itself is not invariant under DSR Lorentz transformations. 
Yet, Refs.~\cite{fins2, fins1} relied crucially on such non-invariant line element for deriving the time dilation, hence the Finsler prediction of type \eqref{GAC2} cannot be a DSR-relativistic result. \vspace{2mm}

This is where the most technical part of the investigations we are here reporting is relevant.
The results on modified time dilation reported in Refs.~\cite{heuristic,fins2, fins1} motivated us to investigate time dilation in the LIV and DSR scenarios, and we quickly established that those results do not apply to the LIV and DSR scenarios. 

As we already observed, in the LIV scenario time dilation is automatically unmodified; therefore, our last task is to investigate time dilation in a DSR scenario with dispersion relation \eqref{GAC1}. 
And for this task the literature provides only partial and/or indirect supporting material.
Several alternative DSR scenarios have been considered,
with or without upholding 
\eqref{GAC1} (see, {\it e.g.}, Refs~\cite{Jafari:2020, carmonaPRD2012}), and the focus of the analyses is typically oriented toward momentum-space features rather 
than spacetime observables.
DSR-relativistic  spacetime observables were studied in Refs~\cite{Mignemi:2005sz,Mignemi:2009ij}, relying however on spacetime coordinates governed by non-standard Poisson brackets, whose relation with standard coordinates suitable for describing the relativistic properties of particles in a laboratory setup remains unclear.
We here investigate time dilation for the most studied DSR scenario (reviewed briefly in the next section), which indeed upholds \eqref{GAC1}, and, in order to allow direct comparison with the results of~\cite{heuristic,fins2, fins1}, we work with spacetime coordinates that satisfy standard ({\it i.e.} trivial) Poisson brackets.
After devoting the next Section to the introduction of the specific DSR scenario on which we base our analysis,
in Section \ref{sec:finiteboosts}, we derive finite boost transformations of all phase-space variables.
In Section \ref{sec:worldlines}, we establish properties of the worldlines that are needed for the derivation of  time dilation.
In Section \ref{sezione metrica}, we determine the DSR-relativistic spacetime metric. In Section \ref{sec:timedilation}, we derive the time dilation factor.
In the closing Section \ref{sec:conclusion} we summarize  our main results and discuss  their phenomenological implications.

\section{The most studied DSR scenario}\label{sec:DSRscenario}
We base our discussion on the most studied DSR model, the one inspired from the structures of the $\kappa$-Poincaré Hopf algebra~\cite{lukierskiFirst,majidRuegg,jurekDSRkPoinc}.
Using the so-called “bicrossproduct basis”, the commutators of the algebra generators read
\begin{equation}\label{AlgebraMomentiBoostRappresentati}
  \begin{aligned}
    &\left[P_\mu,P_\nu\right]=0,
    \quad \left[P_0,N_i\right]=P_i,\\
    \left[P_i,N_j\right]=&-\delta_{ij}\left(\frac{e^{-2 P_0 \ell}-1}{2\ell}-\frac{\ell}{2}\vec{P}^2\right)-\ell P_iP_j,
  \end{aligned}
\end{equation}
where $P_\mu$ are the spacetime translation generators and $N_j$ are the boost generators. The mass Casimir of this algebra is 
\begin{equation}\label{operatore casimir}
C=\frac{4}{\ell^2} \sinh^2{\left(\frac{\ell}{2} P_0\right)}-\vec{P}^2 e^{\ell P_0}.
\end{equation}
These algebra and Casimir describe a deformation of the usual special-relativistic structures, governed by the parameter $\ell$ with dimensions of the inverse of energy. 

We are going to study the effects of such deformations on phase-space variables $\left(x^\mu,p_\mu\right)$ with Poisson brackets
\begin{equation}\label{AlgebraCoordinateMomenti}
    \left\{x^\mu,p_\nu\right\}=-\delta^\mu_\nu, \hspace{5mm}   \left\{x^\mu,x^\nu\right\}=0 , \hspace{5mm}  \left\{p_\mu,p_\nu\right\}=0\,.
\end{equation}
For simplicity, from now on we specialize to the 1+1 dimensional case.\footnote{This allows us to write simpler formulas but does not lead to any loss of generality, as time dilation is essentially a 1+1-dimensional effect.}
The algebra generators and  Casimir can be represented as functions 
on phase space as follows: 
\begin{equation}\label{boost and casimir}
   \begin{aligned}
        P_0=&p_0, \quad 
        N=x^0p_1+x^1\left(\frac{1-e^{-2 \ell p_0}}{2\ell}-\frac{\ell}{2}p_1^2\right),\\
        P_1=&p_1, \quad C=\frac{4}{\ell^2} \sinh^2{\left(\frac{\ell}{2} p_0\right)}-p_1^2 e^{\ell p_0}\,.
   \end{aligned}
\end{equation}

The DSR model introduced here is compatible with the modified dispersion relation of type \eqref{GAC1} and yields a deformed relativistic kinematics.
More specifically, the Hamiltonian constraint $C=m^2$ gives the mass-shell
\begin{equation}\label{mass shell relation}
    \frac{4}{\ell^2} \sinh^2{\left(\frac{\ell}{2} p_0\right)}-p_1^2 e^{\ell p_0}=m^2,
\end{equation}
which reduces to  \eqref{GAC1} at first order in $\ell$. According to this relation, the rest-frame energy $p_0(p_1=0)$ is  $M=\frac{2}{\ell} \sinh^{-1}{(\frac{\ell}{2} m )}$,  coinciding with $m$ up to second order in $\ell$.
Moreover, the Poisson brackets with $C$ and $N$ give, respectively, the dynamical and the boost evolution of any given function $f$ of phase-space variables: $\partial_\tau f=\left\{f, C\right\}$ and $\partial_\eta f=\left\{f, N\right\}$, where $\tau$ is the evolution parameter and $\eta$ is the boost parameter, that is, the rapidity \cite{lukierskiAnnals, jurekDaszkvelocity, mignemiHamiltonianDSR,tamingNonLocality, Barcaroli:2015eqe}. 

From \eqref{AlgebraCoordinateMomenti} and \eqref{boost and casimir} one easily derives the infinitesimal action of the boost on individual phase-space variables:
\begin{equation}\label{poisson brackets di N}
\begin{aligned}
      \left\{x^0,N\right\}&=-e^{-2\ell p_0}x^1,
    \quad \left\{x^1,N\right\}=-x^0+\ell p_1x^1,\\
     \left\{p_0,N\right\}&=p_1, 
    \quad \left\{p_1,N\right\}=\frac{\left(1-e^{-2 p_0 \ell}\right)}{2 \ell}-\frac{\ell}{2} p_1^2,
\end{aligned}
\end{equation}
as well as the infinitesimal action of the Casimir:
\begin{equation}\label{poisson Casimir}
    \begin{aligned}
        \left\{x^0,C\right\}&=\frac{e^{-\ell p_0}+e^{\ell p_0} \left(\ell^2 p_1^2-1\right)}{\ell  },
          \quad \left\{p_0,C\right\}=0,\\ 
         \left\{x^1,C\right\}&=2e^{\ell p_0} p_1,
         \hspace{8mm} \left\{p_1,C\right\}=0.
    \end{aligned}
\end{equation}
To conclude, since the bracket $\left\{p_1,N\right\}$ vanishes on-shell for $p_1=1/|\ell|$, the model predicts a Planckian momentum as second invariant scale beside the speed of light $c$, here set to unity.\footnote{More precisely,  the  invariant speed $c$ is  the velocity of low-energy  photons (with $p_0\ll 1/|\ell|$).}

\section{Finite boost transformation of phase-space variables}\label{sec:finiteboosts}
For our scopes we need the laws of transformation of
phase-space coordinates under finite boosts, that is at
all orders in the rapidity parameter $\eta$. 
Using the Poisson brackets \eqref{poisson brackets di N}, our task  reduces to integrating the following two sets of coupled differential equations
\begin{equation}\label{4xEqs}
   \begin{aligned}
       \partial_\eta p_0 &=
       p_1, 
       \hspace{1.3cm} \partial_\eta p_1 =\frac{\left(1-e^{-2 p_0 \ell}\right)}{2 \ell}-\frac{\ell}{2}p_1^2, \\
       \partial_\eta x^0 &=-e^{-2\ell p_0}x^1,
       \hspace{1cm} \partial_\eta x^1  =-x^0+\ell p_1x^1.
     \end{aligned}
\end{equation}
The first two equations, describing the transformation of momenta, can be directly solved and yield 
\begin{equation}\label{soluzioniMomenti}
    \begin{aligned}
       p_0(\eta)=&\bar{p}_0-\frac{1}{\ell} \ln \frac{4 e^{\eta}}{\left(1+e^{\eta }+\ell \bar{p}_1(e^{\eta}-1)\right)^2-\left(e^{\eta }-1\right)^2 e^{-2\ell\bar{p}_0}},\\
       p_1(\eta)=&\frac{2e^{\eta } \left(\ell^{-1}\sinh\eta \left(\ell^2\bar{p}_1^2-e^{-2 \ell\bar{p}_0}+1\right)+2\bar{p}_1 \cosh\eta\right)}{\left(1+e^{\eta }+\ell\bar{p}_1(e^{\eta}-1)\right)^2-\left(e^{\eta }-1\right)^2 e^{-2\ell\bar{p}_0}},  
    \end{aligned}
\end{equation}
upon using initial conditions $\overline{p}_0= p_0(\eta=0)$ and $\overline{p}_1= p_1(\eta=0)$.

After substituting the solution \eqref{soluzioniMomenti} in the equations \eqref{4xEqs} for spacetime coordinates and setting initial conditions $\overline{x}^0$$=$$ x^0(0)$ and $\overline{x}^1$$=$$ x^1(0)$, we can solve the other two equations in \eqref{4xEqs}, describing the transformation of spacetime coordinates, and find
\begin{equation}\label{soluzioniCoordinate}
    \begin{aligned}
        x^0(\eta)=&\overline{x}^0
                 \frac{\left(1+e^{\eta }+\ell \bar{p}_1(e^{\eta}-1)\right)^2+\left(e^{\eta }-1\right)^2 e^{-2\ell\bar{p}_0}}{\left(1+e^{\eta }+\ell \bar{p}_1(e^{\eta}-1)\right)^2-\left(e^{\eta }-1\right)^2 e^{-2\ell\bar{p}_0}} \\
                 & -\overline{x}^1\frac{2e^{-2\ell\bar{p}_0}\left(1+e^{\eta }+\ell \bar{p}_1(e^{\eta}-1)\right)(e^{\eta}-1)}{\left(1+e^{\eta }+\ell \bar{p}_1(e^{\eta}-1)\right)^2-\left(e^{\eta }-1\right)^2 e^{-2\ell\bar{p}_0}},\\
       x^1(\eta)=&
               \frac{\overline{x}^1}{4} e^{-\eta } \left[\left(1-\overline{p}_1 \ell+e^{\eta }\overline{p}_1 \ell +e^{\eta }\right)^2+\left(e^{\eta }-1\right)^2 e^{-2 \ell \bar{p}_0 }\right] \\
              & -\overline{x}^0 (\sinh\eta +\overline{p}_1 \ell  \cosh \eta-\overline{p}_1 \ell ).
    \end{aligned}
\end{equation}

Equations \eqref{soluzioniMomenti} and \eqref{soluzioniCoordinate} give the finite transformations of phase-space variables under deformed boosts, and they correctly reduce to the standard special-relativistic laws in the limit $\ell\to 0$, as one would expect for a model that analytically deforms special relativity. Furthermore, as expected for a relativistic model, they preserve the form of all Poisson brackets \eqref{AlgebraCoordinateMomenti}, \eqref{poisson brackets di N}, \eqref{poisson Casimir} and of the mass-shell relation \eqref{mass shell relation}, computed using the boosted phase-space variables.

While the transformation of momenta \eqref{soluzioniMomenti} was already known in the DSR literature~\cite{GACandBruno}, the transformation of spacetime coordinates \eqref{soluzioniCoordinate} was missing.
Having filled this gap, we are now able to establish several
properties of the DSR model that are needed for deriving the time dilation effect.

We start by studying their domain of existence.
In special relativity, the rapidity $\eta$ can take any real value.
In this DSR model, instead, such property holds only for $\ell>0$, while for negative $\ell$ the solutions are valid within the finite range $\eta_-$$<$ $\eta<\eta_+$, where
\begin{equation}
        \eta_\pm=\ln{\left(\frac{1\mp e^{\ell \overline{p}_0}\left(\ell\overline{p}_1-1\right)}{1\mp e^{\ell \overline{p}_0}\left(\ell\overline{p}_1+1\right)}\right)}.
\end{equation}
Although the  scale $1/|\ell|$ is invariant for any choice of $\eta$, only for $\ell>0$ this value  also sets the maximum allowed value of spatial momentum: for infinite boost $\eta\rightarrow\pm\infty$ the energy $p_0(\eta)$ diverges but the momentum $p_1(\eta)$ saturates to $1/|\ell|$. 
For $\ell<0$, instead, both quantities diverge as the rapidity approaches the bounds of its domain, $\eta\rightarrow\eta_{\pm}$ \cite{GACandBruno}.\vspace{2mm}

We conclude this section  by showing that  the action on phase space variables  of the inverse of a boost transformation with rapidity $\eta$ is a boost governed by rapidity $\eta'=-\eta$. To do so, it is convenient to write Eqs.~\eqref{soluzioniMomenti}-\eqref{soluzioniCoordinate} in a more compact form as follows.
The transformation of coordinates \eqref{soluzioniCoordinate}, can be cast in the form 
\begin{equation}\label{trasformazione diretta}
    \left(\begin{array}{c} {x^0}  \\{ x^1} \end{array} \right)
    =\Lambda (\bar p_0,\bar p_1;\eta) 
    \left(\begin{array}{c} {\bar x^0}  \\{\bar x^1} \end{array} \right)\,,
\end{equation}
upon appropriate definition of the transformation matrix $\Lambda$. We emphasize again that a major difference
with respect to special relativity is that these laws of transformation of coordinates under finite boosts depend on energy and momentum. Therefore, one needs to specify a point in the full phase-space in order to describe
such transformations.
The boosted energy and momentum variables  are given by the nonlinear functions
\begin{equation}\label{diretta momenti}
   p_0=f_0(\bar p_0,\bar p_1;\eta), \quad
    p_1=f_1(\bar p_0,\bar p_1;\eta),
\end{equation}
whose definition can be straightforwardly read from \eqref{soluzioniMomenti}. Notice that the transformations (\ref{diretta momenti}) can be written  in terms of the matrices $\Lambda$ as follows:
\begin{equation}\label{trasformazioni Lambda momenti}
\frac{\partial}{\partial \bar p_{\nu}}f_{\mu}\left(\bar p_0,\bar p_1;\eta\right)=\left(\Lambda^{-1}\left(\bar p_0,\bar p_1;\eta\right)\right)_{\nu\mu}.    
\end{equation} 

Starting from the transformations \eqref{trasformazione diretta}-\eqref{diretta momenti}, governed by rapidity $\eta$, by virtue of the relativity principle the inverse transformations should take the same form. Namely, for spacetime coordinates one should have
\begin{equation}\label{trasformazione inversa}
    \left(\begin{array}{c} {\bar x^0}  \\{\bar x^1} \end{array} \right)
    =\Lambda (p_0,p_1;\eta') 
    \left(\begin{array}{c} {x^0}  \\{ x^1} \end{array} \right)\,,
\end{equation}
and for momenta
\begin{equation} \bar p_0=f_0(p_0,p_1;\eta'),\quad \bar p_1=f_1(p_0,p_1;\eta')\,,
\end{equation}
for some $\eta'(\eta)$.
Using again Eqs.~\eqref{trasformazione diretta}-\eqref{diretta momenti} to write the variables $(x^0,x^1,p_0,p_1)$ appearing in the above equations one finds the relations
\small
\begin{equation}\label{inverse inverse}
    \left(\begin{array}{c} {\bar x^0}  \\{ \bar x^1} \end{array} \right)
    =\Lambda \left(f_0(\bar p_0,\bar p_1;\eta),f_1(\bar p_0,\bar p_1;\eta);\eta'\right) 
   \Lambda (\bar p_0,\bar p_1;\eta) 
    \left(\begin{array}{c} {\bar x^0}  \\{\bar x^1} \end{array} \right)\,,
\end{equation}
\normalsize
for coordinates and
\begin{equation} \begin{aligned}
\bar p_0&=f_0\left(f_0(\bar p_0,\bar p_1;\eta),f_1(\bar p_0,\bar p_1;\eta);\eta'\right),\\
\bar p_1&=f_1\left(f_0(\bar p_0,\bar p_1;\eta),f_1(\bar p_0,\bar p_1;\eta);\eta'\right)\,,
\end{aligned}
\end{equation}
for momenta.
A simple computation shows that the only solution of these equations is  $\eta'=-\eta$.

\section{Worldline covariance}\label{sec:worldlines}

Throughout the rest of the manuscript, we will be concerned with the relativistic kinematics of free particles.
Let us consider a free particle with rest energy $M$, energy $\epsilon$, spatial momentum $p$ and spacetime coordinates $(t,x)$. The particle is on-shell, so that energy and momentum satisfy the relation \eqref{mass shell relation} and its spacetime coordinates lie on a worldline.
This can be derived by integrating the Poisson brackets \eqref{poisson Casimir} to solve Hamilton's equations $\partial_\tau t=\left\{t, C\right\}$ and $\partial_\tau x=\left\{x, C\right\}$ and reads:
\begin{equation}\label{genericWLoff-shell}
    x_{\epsilon,p}(t)=V(\epsilon,p)\hspace{1mm}t+x(0),
\end{equation}
where $x(0)\equiv x_{\epsilon,p}(t=0)$ 
and
\begin{equation}
\label{velocità}
V(\epsilon,p) = \frac{2\ell p e^{2\ell\epsilon}}{e^{2\ell\epsilon}\left(\ell^2p^2-1\right)+1},
\end{equation}
is the coordinate velocity\footnote{Since translations act trivially on the spacetime coordinates defined by \eqref{AlgebraCoordinateMomenti}, the coordinate velocity coincides with the physical particle velocity \cite{kbob,jackMignemiPhysVel}.}\textsuperscript{,}\footnote{For later ease of notation we are not making the on-shell condition explicit when writing the velocity.  
Once this is accounted for, the velocity reads 
$
V(\epsilon) =\text{sign}(p)\frac{\sqrt{2} e^{\frac{3 \ell\epsilon}{2}} \sqrt{\cosh (\ell\epsilon)-\cosh (\ell M)}}{\text{sign}(\ell)\left(1-e^{\ell \epsilon} \cosh (\ell M)\right)}$. } $V\equiv\frac{dx}{dt}=\frac{\left\{x, C\right\}}{\left\{t, C\right\}}$
\cite{jurekDaszkvelocity,mignemiVelocity,kbob,jackMignemiPhysVel}.

Relativistic consistency requires that this worldline be covariant under deformed boosts.
This has been already shown to hold when considering infinitesimal boosts (first order in $\eta$) \cite{tamingNonLocality, Giulia+Sjors}.
Using the finite boost transformations \eqref{trasformazione diretta}-\eqref{diretta momenti}, that transform $(\epsilon,p,t,x)$ to $(\epsilon',p',t',x')$ according to
\begin{equation}\label{trasformazione diretta particella}
    \left(\begin{array}{c} {t'}  \\{ x'} \end{array} \right)
    =\Lambda (\epsilon,p;\eta) 
    \left(\begin{array}{c} {t}  \\{x} \end{array} \right)\,,
\end{equation}
and
\begin{equation}\label{diretta momenti particella}
   \epsilon'=f_0(\epsilon,p;\eta), \quad
    p'=f_1(\epsilon,p;\eta),
\end{equation}
we are now able to prove worldlines covariance under finite boosts.

We use the worldline relation \eqref{genericWLoff-shell} to substitute $x$ with $x_{\epsilon,p}(t)$ in \eqref{trasformazione diretta particella}, and combine the resulting two relations linking $t'$ and $x'$ to $t$, so that we find:
\begin{equation}\label{worldline nel sistema primato off-shell}
    x'(t')=\frac{\Lambda_{10}+V\Lambda_{11}}{\Lambda_{00}+V\Lambda_{01}} t' + \left[\Lambda_{11}-\Lambda_{01}\left(\frac{\Lambda_{10}+V\Lambda_{11}}{\Lambda_{00}+V\Lambda_{01}}\right)\right] x(0),
\end{equation}
where $\Lambda_{\mu\nu}$ are the matrix elements of $\Lambda(\epsilon,p;\eta)$ and $V=V(\epsilon, p)$ is the coordinate velocity \eqref{velocità}.
To prove covariance, we  show that Eq.~\eqref{worldline nel sistema primato off-shell} can be cast in the same functional form as the worldline \eqref{genericWLoff-shell} written for primed coordinates, that is
\begin{equation}\label{worldline primata}
     x'_{\epsilon',p'}(t')=V(\epsilon',p')\hspace{1mm}t'+{x}'(0),
\end{equation}
where $x'(0)\equiv x'_{\epsilon',p'}(t'$$=$$0)$.
Concerning velocity, a simple computation shows that $V'=V(\epsilon',p')$ where $V'\equiv\frac{dx'}{dt'}$ is read from Eq.~\eqref{worldline nel sistema primato off-shell} and the functional form of $V(\epsilon',p')$ is the one of Eq.~\eqref{velocità}, with $\epsilon'$ and $p'$ given by Eq.~\eqref{diretta momenti particella}. Notice that  
\begin{equation}\label{velocity composition}
    V' = \frac{\Lambda_{10}+V\Lambda_{11}}{\Lambda_{00}+V\Lambda_{01}},
\end{equation}
encodes the (deformed) rule of composition of velocities.

Concerning the initial condition in Eq.~\eqref{worldline primata},
\begin{equation}
    {x}'(0)  =x'_{\epsilon',p'}(t')-V(\epsilon',p')\hspace{1mm}t',
\end{equation}
we use Eq.~\eqref{trasformazione diretta particella} to write primed spacetime coordinates and Eq.~\eqref{velocity composition} to write the  velocity $V(\epsilon',p')=V'$, obtaining:
\begin{equation}
  {x}'(0)  =\Lambda_{10}t+\Lambda_{11}x-\frac{\Lambda_{10}+V\Lambda_{11}}{\Lambda_{00}+V\Lambda_{01}} \left( \Lambda_{00}t+\Lambda_{01}x\right)\,. 
\end{equation}
Further asking that the worldline relation \eqref{genericWLoff-shell} holds for $(t,x)$, one finds
\begin{equation}
\label{trasformazione delle intercette}
x'(0) = \left(\Lambda_{11}-\frac{\Lambda_{10}+V\Lambda_{11}}{\Lambda_{00}+V\Lambda_{01}}\Lambda_{01}\right)x\left(0\right) \,,
\end{equation}
which is exactly the time-independent term of Eq.~\eqref{worldline nel sistema primato off-shell}.
We then conclude that the worldlines of free particles are fully covariant under boost transformations.

The results obtained in this Section so far allow us to find the relation between a particle's velocity $V$ and rapidity $\eta$.
In special relativity, they are connected by the familiar relation $V=\tanh \eta$.
In DSR, this standard relation is deformed. To see this, let us consider a particle at rest, so that $\epsilon'=M$, $p'=0$, and  $V'=0$. 
After a boost with rapidity $-\eta$, the particle's velocity is given by  \eqref{velocity composition}, which in this case reads $V=\Lambda_{10}(M,0;-\eta)/\Lambda_{00}(M,0;-\eta)$.
Writing the matrix elements of $\Lambda(M,0;-\eta)$ explicitly we find the deformed relation
\begin{equation}\label{velocity boosted}
    V=\frac{\sinh (\eta) [\cosh (\eta) \sinh (\ell  M )+\cosh (\ell  M)]}{\cosh (\eta ) \cosh (\ell  M )+\sinh (\ell  M)}\,,
\end{equation}
which correctly reduces to the special-relativistic formula for $\ell \rightarrow 0$.

\section{Covariant spacetime metric}
\label{sezione metrica}
It is  well established that the  mass-shell relation of a DSR theory defines a metric on momentum space, such that the mass-shell relation is given by the geodesic distance of the point $(\epsilon,p)$ from the origin  \cite{GACrelativelocality,Giulia+Flaviorelativelocality,diffeo}.\footnote{Other properties of DSR models, such as the laws of addition of energy and momentum, define torsion and nonmetricity \cite{Giulia+Flaviorelativelocality,trevisan}}
Because of the relativistic properties of DSR models, these possibly curved momentum space metrics are maximally symmetric,   that is they allow for a full set of relativistic generators. 
For instance, the momentum space metric of the  DSR model we are considering here is known to be that of a de Sitter manifold \cite{jurekDSRdeSitter1,jurekDSRdeSitter2,Giulia+Flaviorelativelocality,diffeo,trevisan,Amelino-Camelia:2013sba, Amelino-Camelia:2013uya}:
 the mass-shell relation \eqref{mass shell relation} can be obtained  from the metric $\Tilde{g}(\epsilon)=\text{diag}(1,-e^{2\ell \epsilon})$, which is indeed a de Sitter metric on momentum space.\footnote{An example of anti-de Sitter momentum spaces can be found in \cite{Arzano:2014jua}.} 

Having established the laws of transformation  of  phase-space coordinates under finite boost, we are now in the position of showing that 
the inverse of the momentum space metric, 
$g(\epsilon)=\text{diag}(1,-e^{-2\ell \epsilon})$, is the covariant metric on spacetime at all orders in $\ell$, as  seen by a free particle with energy $\epsilon$ and momentum $p$, obeying the mass-shell relation \eqref{mass shell relation}.
The corresponding line element 
\begin{equation}\label{metric}
    ds^2=dt^2-e^{-2\ell \epsilon}dx^2,
\end{equation}
is in fact  invariant under the finite boost transformations \eqref{trasformazione diretta}-\eqref{diretta momenti}, which take the form \eqref{trasformazione diretta particella}-\eqref{diretta momenti particella} when applied to on-shell variables.
To see this, we start from the line element written in terms of boosted coordinates, $ds'^2= dt'^2-e^{-2\ell \epsilon'}dx'^2$, and use  transformations \eqref{trasformazione diretta particella}-\eqref{diretta momenti particella} to  rewrite it in terms of the old variables.
Eq.~\eqref{diretta momenti particella} gives the boosted energy $\epsilon'=f_{0}(\epsilon,p;\eta)$, while the differentials $dt'$ and $dx'$ can be written as
\begin{equation}\label{differetials}
    \begin{aligned}
        dt'=&\partial_t t' dt+\partial_x t'dx+\partial_\epsilon t' d\epsilon+\partial_p t'dp\,,\\
         dx'=&\partial_t x' dt+\partial_x x'dx+\partial_\epsilon x' d\epsilon+\partial_p x'dp\,,
    \end{aligned}
\end{equation}
where the last two terms of both equations vanish since $d\epsilon=dp=0$ for free particles. The derivatives multiplying $dt$ and $dx$ correspond to the components of the transformation matrix $\Lambda(\epsilon,p;\eta)$ of Eq.~\eqref{trasformazione diretta particella}.
Therefore, we obtain
\begin{equation}\label{boosted line element}
    ds'^2= (\Lambda_{00}dt+\Lambda_{01}dx)^2-e^{-2\ell f_0}(\Lambda_{10}dt+\Lambda_{11}dx)^2\,.
\end{equation}
A direct computation shows that the dependence on $\eta$ contained in the matrix elements $\Lambda_{\mu\nu}$ cancels out and the boosted line element $ds'^2$ in \eqref{boosted line element} coincides with the line element $ds^2$.
We then conclude that the metric defined in \eqref{metric} is covariant under finite boosts.
This is in agreement with previous results obtained at  first order in $\ell$~\cite{NiccoMetric,gacGiuliaLiberati}.\vspace{2mm}

In contrast,  another frequently-used candidate spacetime metric $h(\epsilon,p)$, denoted  rainbow metric and defined asking that its inverse $\tilde{h}(\epsilon,p)$ reproduces the mass-shell (Hamiltonian) relation through
$\sum_{\mu\nu}\tilde{h}_{\mu\nu}(\epsilon,p) p_\mu p_\nu = m^2$ \cite{Magueijo:2002xx, Letizia:2016lew, Lobo:2018fym}, is not covariant. In fact, 
following the same steps outlined above, one can easily show that the line element induced by the rainbow metric 
$h=\text{diag}( \frac{\ell^2 \epsilon^2}{4} \sinh^{-2}{(\frac{\ell}{2} \epsilon)},-e^{-\ell \epsilon})\approx\text{diag}(1,-1+\ell \epsilon)$, obtained from the mass-shell \eqref{mass shell relation},  is not invariant under deformed boosts \eqref{trasformazione diretta particella}-\eqref{diretta momenti particella}, already at the first order in the deformation parameter $\ell$.
This kind of rainbow metric is the one commonly used 
to obtain the Finsler line element \cite{fins1, fins2}. 
And indeed  the Finsler line element is not invariant under deformed boosts, as was proven  in~\cite{mignemiFinslerNoDSR, gacGiuliaLiberati}.\footnote{Remarkably, in~\cite{gacGiuliaLiberati} it was  shown that the Finsler framework can be adapted to a DSR-relativistic scenario by adding a suitable boundary term to the Finsler Lagrangian, so that the  line element corresponds to the one induced by the metric $g(\epsilon)=\text{diag}(1,-e^{-2\ell \epsilon})\approx\text{diag}(1,-1+2\ell \epsilon)$.}
This explains why the time dilation of Eq.~\eqref{GAC2} proposed in \cite{fins1} is incompatible with a DSR-relativistic scenario.\vspace{2mm}

We conclude this section by outlining a few interesting side results which follow immediately from our understanding of the covariant spacetime metric $g(\epsilon)$.

Using the spacetime metric, in special relativity one can define a finite spacetime interval that is invariant under Lorentz transformation (but not translations), namely $I_{SR}=t^2-x^2$.
Likewise, we find that also in our DSR setting there exists a similar relativistic scalar,
\begin{equation}
     I_{DSR}=t^2-e^{-2\ell \epsilon}x^2,
\end{equation}
that can be easily verified to be invariant under the deformed boost transformations \eqref{trasformazione diretta particella}-\eqref{diretta momenti particella}.

Moreover,  the covariance of $g$ can be written in a compact form using the matrix notation:
\begin{equation}
\label{covarianza metrica}
g(\epsilon') = \Lambda^T(\epsilon,p;\eta)\hspace{1mm} g(\epsilon) \hspace{1mm}\Lambda(\epsilon,p;\eta).
\end{equation}
This relation implies the invariance of the line element $ds^2$ and  can be understood as a DSR generalization of the defining property of  standard Lorentz matrices.
It also implies the invariance of the (de Sitter) line element in momentum space: 
$\sum_{\mu\nu}\tilde{g}_{\mu\nu}\left(\epsilon'\right)dp_{\mu}'dp_{\nu}'=\sum_{\mu\nu}\tilde{g}_{\mu\nu}\left(\epsilon\right)dp_{\mu}dp_{\nu}$.
This allows to understand Eqs.~(\ref{diretta momenti}) and~(\ref{trasformazioni Lambda momenti}) in geometric terms: they describe   the transformation laws of momenta as those of (comoving) coordinates of de Sitter momentum  space.

\section{Time Dilation}\label{sec:timedilation}
 
In special relativity, the proper time interval $\Delta \tau$ of a particle and the corresponding time interval $\Delta t$ measured by a boosted observer are related by $\Delta t=\gamma_{sr}\Delta\tau$, where the time dilation factor is $\gamma_{sr}$$=\cosh{\eta}=\epsilon/M$, written in terms of rapidity $\eta$ or of energy $\epsilon$ of the particle, respectively. This special-relativistic result can be equivalently obtained  either from the Lorentz transformation of the time interval or from the Minkowski metric. In this Section, we  use the relations derived so far  to compute the time dilation factor in the DSR model, showing that the two methods give equivalent (but deformed) results also in the DSR scenario.\vspace{2mm}

In a particle's rest frame its worldline reads $x'_{\epsilon'=M,p'=0}(t')=0$, where $t'$ is the particle's proper time $\tau$. Applying the transformations \eqref{trasformazione inversa} to the particle's (proper) time interval $\Delta t'\equiv\Delta \tau$ and accounting for the worldline relation $x'(t')=0$, one finds that the  boosted time interval $\Delta t$ is related to $\Delta\tau$ via the proportionality factor  $\Lambda_{00}(M,0;-\eta)$, that is: 
\begin{equation}\label{time dilation formula}
    \Delta t =\frac{\cosh (\eta )  \cosh (\ell M )+\sinh (\ell M )}{\cosh (\ell M )+\cosh (\eta ) \sinh (\ell M )} \Delta\tau.
\end{equation}

One can derive the same relation by using the invariance of the line elements $ds^{2}=ds'^2$, which can be recast in the form 
\begin{equation}\label{dilation from metric}
           dt\sqrt{1-e^{-2\epsilon \ell}{V}^2}=d\tau,
\end{equation}
where we  defined the coordinate velocity $V$$\equiv$$dx/dt$ as in Section~\ref{sec:worldlines} and we used the fact that $ds'$$=$$d\tau$ in the particle's rest frame.
Using \eqref{diretta momenti} and \eqref{velocity boosted} with $\eta'=-\eta$ to write $\epsilon$ and $V$ in terms of the rest-frame quantities $(\epsilon'$$=$$M,p'$$=$$0)$, we can integrate Eq.~\eqref{dilation from metric} to find the same relation 
as in Eq.~\eqref{time dilation formula}.
Therefore, in our DSR model one has $\Delta t= \gamma_{dsr}\Delta \tau$, with 
\begin{equation}\label{gamma dsr}
    \gamma_{dsr}= \frac{\cosh (\eta )  \cosh (\ell M )+\sinh (\ell M )}{\cosh (\ell M )+\cosh (\eta ) \sinh (\ell M )}\,.
\end{equation}

The time dilation factor $\gamma_{dsr}$ can also be written in terms of the particle's energy rather than rapidity.  Substituting  the general expression for the velocity \eqref{velocità} taken on-shell into Eq. \eqref{dilation from metric} one obtains
\begin{equation}\label{time dilation formula in energy}
    \Delta t =\frac{\cosh (\ell M)-e^{-\ell \epsilon}}{\sinh (\ell M)} \Delta\tau,
\end{equation}
which is equivalent to \eqref{time dilation formula}, as can be  seen by using the transformation \eqref{diretta momenti particella} that connects $\epsilon$ to $M$ and $\eta$.
Therefore, we can also write 
\begin{equation}\label{gamma dsr energy}
    \gamma_{dsr}= \frac{\cosh (\ell M)-e^{-\ell \epsilon}}{\sinh (\ell M)}\,.
\end{equation}

The time dilation factor $\gamma_{dsr}$   reduces to the standard factor $\gamma_{sr}$$=\cosh{\eta}=\epsilon/M$ in the undeformed limit $\ell\to 0$ and is fully analytical in the deformation parameter $\ell$, in contrast to previous proposals described in Section~\ref{sec:preliminaries}.
Moreover, $\gamma_{dsr}$ is well-defined, positive and monotonous for all possible values of $M$ and $\epsilon$ (or $\eta$), that is, it has no pathologies and always preserves causality.
Looking at the infinite-boost limit, when $\ell>0$ the time dilation factor $\gamma_{dsr}$ saturates to the finite value $\coth{(\ell M)}$ for $\eta\to\pm\infty$, while for negative deformation parameter $\ell<0$ the infinite boost limit corresponds to $\eta\to\eta_{\pm}$ and the factor $\gamma_{dsr}$ is  divergent.\footnote{The infinite-boost limit can be equivalently taken in terms of energy:\ for both signs of $\ell$ it reads $\epsilon\to\infty$ and one finds the same saturation property of $\gamma_{dsr}$, as can be directly checked from \eqref{gamma dsr energy}.}
This can be compared to special relativity, where the time dilation factor $\gamma_{sr}$ diverges for $\eta\to\pm\infty$.

Our result allows us to clarify the conceptually different roles played by the gamma factor in special relativity, which lead to nonequivalent gamma factors in DSR models.
In special relativity,  the Lorentz factor $\gamma_{sr}$ associated to a particle of mass $M$ and energy $\epsilon$  can be equivalently computed as 
\begin{equation}\label{eq:gammaSR}
\gamma_{sr}=\frac{\Delta t}{\Delta\tau} =\frac{\epsilon}{M}=\cosh{\eta}\,.
\end{equation}
In DSR these quantities are   different:   from time dilation one finds the gamma factor of Eq.~\eqref{gamma dsr} (or, equivalently, Eq.~\eqref{gamma dsr energy}), 
while using Eq.~\eqref{diretta momenti particella} to compute the ratio between the particle's energy and mass one finds 
\begin{equation}
\tilde\gamma_{dsr}\equiv \epsilon/M = \frac{\ln \left(\cosh (\eta ) \sinh (\ell  M)+\cosh (\ell  M)\right)}{\ell  M}\,.
\end{equation}

\section{Summary and Outlook}\label{sec:conclusion}
In this paper, we investigated the time dilation effect within theories characterized by a Planck-scale modified dispersion relation of the form \eqref{GAC1}.
Having observed that previous proposals, described by Eq.~\eqref{GAC2} and derived independently in \cite{heuristic} and \cite{fins1}, are consistent neither with standard Lorentz-invariance violation scenarios nor with DSR scenarios, we derived the time dilation effect within  a fully DSR-relativistic model compatible with the dispersion relation \eqref{GAC1}, and adopting standard ({\it i.e.}\ trivial)
Poisson brackets among
spacetime coordinates.\vspace{2mm}

We derived the finite boost transformations on all phase-space variables and to all orders in the deformation parameter.
We showed that free particle  worldlines are fully covariant under these boost transformations, generalizing previous results obtained at first order in the deformation parameter, and we identified the covariant spacetime metric. 
These results provide the ingredients needed to compute the time dilation effect, encoded in Eq.\eqref{time dilation formula in energy}.
Having obtained a formula valid to all orders in the deformation parameter, we could verify that the associated dilation factor $\gamma_{dsr}$ is well-defined and monotonous for all boosts.
This is in contrast with other descriptions of time dilation beyond special relativity, encoded in Eq.~\eqref{GAC2}, as discussed in Section \ref{sec:preliminaries}.
This qualitative difference translates into different phenomenological implications.

\begin{figure}[htbp]
  \centering
  \includegraphics[width=0.95\linewidth]{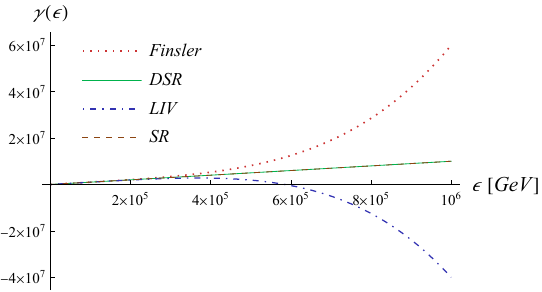}
  \caption{\footnotesize Time dilation factors as a function of energy for high-energy muons ($\epsilon \gg M \simeq 100\,\text{MeV}$). The deformation scale is taken to be Planckian ($\ell^{-1}\simeq10^{19}\,\text{GeV}$). The heuristic LIV prediction, Eq.\ \eqref{GAC2}, and the Finsler prediction, Eq.\ \eqref{GAC2} with opposite sign, are compared to our DSR prediction in Eq.\ \eqref{gamma dsr energy} and to the standard special relativity (SR) result, Eq.~\eqref{eq:gammaSR}. Both the LIV and Finsler curves deviate significantly from the special-relativistic curve already at relatively low energies around $10^6\,\text{GeV}$. In contrast, the DSR curve essentially overlaps with the special-relativistic one throughout this low-energy range.}
  \label{fig:plot Finsler/Auger vs DSR}
\end{figure}

\begin{figure}[htbp]
  \centering
  \includegraphics[width=0.95\linewidth]{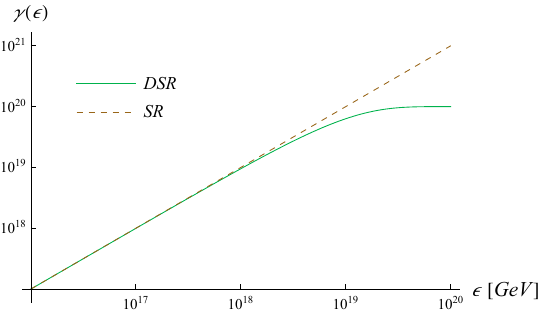}
  \caption{\footnotesize Time dilation factors as a function of energy for high-energy muons ($\epsilon \gg M \simeq 100\,\text{MeV}$). As in Fig.\ \ref{fig:plot Finsler/Auger vs DSR}, we take $\ell^{-1}\simeq10^{19}\,\text{GeV}$. The DSR prediction in Eq.\ \eqref{gamma dsr energy} starts to deviate significantly from the special-relativistic (SR) one only at high energies around $10^{18}\,\text{GeV}$.}
  \label{fig:plot SR vs DSR}
\end{figure}

For a high-energy particle with energy $\epsilon$ and mass $M$, such that $|\ell|^{-1}\gg \epsilon\gg M$, our formula \eqref{time dilation formula in energy} predicts a time dilation factor
\begin{equation}\label{first order}
    \gamma_{dsr}(\epsilon) \approx \frac{\epsilon}{M}\left[1 -\frac{\ell \epsilon
    }{2}\right]\,.
\end{equation}
Not surprisingly, the correction with respect to the standard factor is governed by $\ell \epsilon$, a quantity that is small as long as the energy is subplanckian.  
In other words, our fully relativistic DSR derivation provides an analytical deformation of the standard time dilation formula.
Because of this, phenomenological effects that were estimated to be significant for the scenarios leading to Eq.~\eqref{GAC2} are much smaller in this scenario (see Fig.~\ref{fig:plot Finsler/Auger vs DSR} and Fig.~\ref{fig:plot SR vs DSR}).

For example, for an atmospheric muon with primary energy $10^{18}$ eV, as considered in Ref.~\cite{heuristic}, the relative difference between the time dilation factor \eqref{GAC2} and the special-relativistic one is an impressive $10^{9}$, while using the DSR time dilation factor \eqref{first order} the relative difference with respect to the special-relativistic case is just $10^{-11}$.

The two models have significantly different implications also for $1$ TeV LHC muons, considered in Ref.~\cite{fins1}. In this case the relative difference of the values of \eqref{GAC2} and \eqref{first order} with respect to the special-relativistic time dilation are $10^{-9}$ and $10^{-17}$, respectively. And in fact, while Ref.~\cite{fins1} estimated that LHC muon lifetime measurements allow us to set a lower bound on the deformation parameter appearing in \eqref{GAC2} which is just short of three orders of magnitude from the Planck scale, $|\ell|^{-1}\gtrsim  10^{16}$ GeV, by following a similar line of reasoning one would find for the deformation parameter in Eq.~\eqref{first order} a lower bound that is several orders of magnitude away from the Planck-scale, $|\ell|^{-1}\gtrsim 10^9$ GeV.
It therefore appears that, unlike other DSR-relativistic predictions which instead are motivating a lively phenomenological program~\cite{gacLivingREview,COSTreview}, DSR predictions for time dilation might not be testable in the near future.

Evidently, a natural next step for this research direction will be to verify whether the features we found in the specific DSR scenario here adopted are also present in other DSR scenarios.
We conjecture that most of our results will be of general applicability, since in deriving them we developed the intuition that we were dealing with features connected primarily with just the presence of relativistic invariance, rather than with specific details of the DSR scenario.\vspace{2mm}

\begin{acknowledgments}
P.P.\ acknowledges Lucio Vacchiano and Francesco Giovinetti for helpful discussions.
This work falls within the scopes of the  COST Action CA23130 “Bridging high and low energies in search of quantum gravity”.
\end{acknowledgments}

\end{document}